\documentclass[prb,amsmath,amsfonts,amssymb,numbers,twocolumn,floatfix]{revtex4}
\usepackage{graphicx}
\usepackage{bm}
\usepackage{color}

\newcommand{\braket}[2]{\left\langle \, #1 \, | \, #2 \right \rangle}

\newcommand{\A}{\mathbf{A}}
\newcommand{\B}{\mathbf{B}}
\newcommand{\T}{\mathbf{T}}
\newcommand{\X}{\mathbf{X}}
\newcommand{\Y}{\mathbf{Y}}
\newcommand{\w}{\bm{\omega}}
\newcommand{\J}{\mathcal{J}}
\newcommand{\K}{\mathcal{K}}
\newcommand{\Q}{\mathbf{Q}}
\newcommand{\erfc}{\mathrm{erfc}}
\newcommand{\erf}{\mathrm{erf}}
\newcommand{\GDV}{\textsc{gaussian}}

\newcommand{\LRHF}{{LC-$\omega$LDA}}
\newcommand{\LRdRPA}{{LC-$\omega$LDA+dRPA}}

\begin{document}
\bibliographystyle{apsrev}

\title{Long-range-corrected hybrids including RPA correlation}
\author{Benjamin G. Janesko, Thomas M. Henderson, and Gustavo E. Scuseria}
\affiliation{Department of Chemistry, Rice University \\ Houston, Texas, USA 77005 
}

\begin{abstract}
We recently demonstrated a connection between the random phase approximation
(RPA) and coupled cluster theory [J.  Chem. Phys. {\bf{129}}, 231101 (2008)].
Based on this result, we here propose and test a simple scheme for introducing
long-range RPA correlation into density functional theory. Our method
provides good thermochemical results and models van der Waals interactions
accurately.
\end{abstract}

\maketitle

We recently demonstrated that the ground state correlation energy associated
with the random phase approximation
(RPA)\cite{Oddershede,RingSchuck,Furche2001,Kresse,Dobson2005,Furche2005,Furche2008b,Scuseria2009}
is connected with an approximate ring coupled cluster doubles (rCCD) approach. 
The RPA excitation problem requires the solution of
\begin{equation}
\begin{pmatrix}
\hfill \A & \hfill \B \\  -\B & -\A
\end{pmatrix}
\begin{pmatrix}
\X \\ \Y
\end{pmatrix}
=
\begin{pmatrix}
\X \\ \Y
\end{pmatrix}
\w,
\label{RPA}
\end{equation}
where the matrices $\A$, $\B$, $\X$, and $\Y$ are of dimension 
$o v \times o v$, with $o$ and $v$ being the number of occupied and unoccupied 
spin-orbitals, respectively.  The plasmonic formula\cite{RingSchuck} for the 
RPA ground state correlation energy is
\begin{equation}
E_c^{RPA} =  \frac{1}{2} \mathrm{Tr}(\w - \A).
\label{EcDRPA}
\end{equation}
As shown in our previous paper,\cite{Scuseria2009} Eq. \ref{RPA} is equivalent 
to 
\begin{equation}
\B + \A \, \T + \T \, \A + \T \, \B \, \T = \bm{0},
\label{Riccati}
\end{equation}
an approximate coupled cluster (CC) doubles equation with excitation amplitudes
$\T = \Y \, \X^{-1}$.  Further, the RPA correlation energy can be evaluated 
from the CC-like expression
\begin{equation}
E_c^{rCCD} = \frac{1}{2} \mathrm{Tr}(\B \, \T) = \frac{1}{2} \mathrm{Tr}(\w - \A).
\label{EcrCCD}
\end{equation}

The excitation amplitudes $\T$ imply the existence of an underlying wave
function in RPA. This lets us follow Savin, Stoll, and
coworkers\cite{Stoll1985,rsh1} and introduce range separation between density
functional theory (DFT) and wave function theory, with RPA for the long-range
correlation.  Range separation is a powerful
technique\cite{Leininger1997,Iikura2001,Heyd2003,Toulouse2004,Goll2005,Angyan,Goll2006,Gerber2007,Henderson2007}
that can improve upon both standard wave function methods and semilocal
exchange-correlation (xc) functionals.  Range separated DFT partitions the
electron-electron interaction operator into short (SR) and long (LR) ranges 
\begin{equation}
\label{eq:rsh}
\frac{1}{r_{12}} = \underbrace{\frac{\erfc(\w r_{12})}{r_{12}}}_{SR} 
                 + \underbrace{\frac{\erf(\w  r_{12})}{r_{12}}}_{LR} , 
\end{equation}
and typically (but not always, see Refs. \onlinecite{Heyd2003,Henderson2007}) treats the SR
(LR) component with semilocal (wave function) approximations. 

Toulouse \textit{et al.}\cite{Toulouse2009} recently proposed a range-separated
treatment of RPA via the adiabatic-connection fluctuation-dissipation
theorem,\cite{Kohn1998,Furche2001,Furche2005}  combining a short-range
semilocal xc functional with long-range full RPA.  Consistent with other
work,\cite{Goll2006,Gerber2007} they find that their long-range correlation
energy has a relatively weak basis set dependence.  They also remove some
artifacts of full-range RPA, including a "bump" in the symmetry-restricted
singlet Be$_2$ dissociation curve.

{\em Theory.} 
We propose a simpler long-range RPA based on the connection to coupled
cluster theory discussed above, which avoids a costly adiabatic connection integral.  
We evaluate long-range RPA as a
one-shot correction to a self-consistent generalized Kohn-Sham\cite{Seidl1996} 
(GKS) calculation combining long-range exact (Hartree-Fock-type, HF) exchange 
and short-range local spin density (LSDA) exchange-correlation. Our xc energy 
is 
\begin{equation}
\label{eq:Exctot}
E_{xc} = E_{xc}^{SR-LSDA} + E_x^{LR-HF} + c_{RPA}\ E_c^{LR-RPA}.
\end{equation}
We evaluate $E_c^{LR-RPA}$ with Eq. \ref{Riccati}-\ref{EcrCCD}, and build $\A$
and $\B$ from the long-range two-electron integrals and the GKS spin-orbitals
and orbital energies. The coefficient $c_{RPA}$ is discussed below.

In this work, we focus on what we will refer to as direct RPA.   In 
the (real) canonical spin-orbital basis we use throughout this paper, the 
direct RPA matrices are
\begin{subequations}
\begin{alignat}{1}
A_{ia,jb} &= (\epsilon_a - \epsilon_i) \delta_{ij} \delta_{ab} + \braket{ib}{aj},
\\
B_{ia,jb} &= \braket{ij}{ab}.
\end{alignat}
\label{DefAB}
\end{subequations}
Here $\epsilon$ is a generalized Kohn-Sham orbital energy.  Indices $i$ and
$j$ indicate occupied spin-orbitals, $a$ and $b$ indicate virtual
spin-orbitals, and $\braket{ij}{ab}$ is a two-electron integral in Dirac's
notation.  For real orbitals, note that $\braket{ib}{aj}=\braket{ij}{ab}$. What we refer to as full RPA uses antisymmetrized
two-electron integrals in Eq. \ref{DefAB}.  

For practical calculations, direct RPA has the great advantage that the
correlation energy is guaranteed to be real if the orbitals obey the aufbau
principle.  This is not true of full RPA, where instabilities in the reference
determinant\cite{Seeger1977} can produce a complex correlation
energy.\cite{Furche2001} In cases such as Be$_2$ where the reference has
triplet instabilities,\cite{Malrieu} one may have to limit full RPA to singlet
excitations, while with direct RPA no such restriction is needed. 

An additional advantage of direct RPA is that it reduces the dimension of the
problem compared to full RPA (ring CCD with antisymmetrized two-electron
integrals).  Consider the block of $\B$ corresponding to spatial orbitals
$\varphi_I$, $\varphi_A$, $\varphi_J$, and $\varphi_B$ and spin ordering
$\alpha \alpha$, $\beta \beta$, $\alpha \beta$, and $\beta \alpha$.  The $\B$
matrix for full RPA and closed shells becomes 
\begin{equation}
\B_{IA,JB} = 
\begin{pmatrix}
\J -\K & \J & 0 & 0 \\
\J & \J -\K  & 0 & 0 \\
 0 &  0 & 0 & -\K \\
 0 &  0 & -\K & 0
\end{pmatrix},
\label{eq:Bdecompose}
\end{equation}
with $\J = \braket{IJ}{AB}$ and $\K = \braket{IJ}{BA}$.  For open shells 
(unrestricted), the $\J$ and $\K$ entries in different blocks will generally 
differ.  Direct RPA ($\K=0$) zeros the ``spin-flip'' block of $\B$ containing 
only $-\K$, reducing the dimension by a factor of 2 compared to full RPA.  For 
both closed and open shell systems, we can diagonalize the remaining upper 
block of $\B$ with the unitary transformation 
\begin{equation} 
\Q = \frac{1}{\sqrt{2}}
\begin{pmatrix}
1 & \hfill 1 \\ 1 & -1  
\end{pmatrix}.
\label{eq:Q}
\end{equation}
For closed shells, $\Q$ is a spin-adaptation operator. The resulting $2\times
2$ diagonal matrix for direct RPA has only one non-zero eigenvalue,
corresponding to singlet excitations. Since only the singlet block of $\B$ is
non-zero, we only need the singlet part of $\T$ to evaluate the correlation
energy of Eq.  \ref{EcrCCD}. Additionally, in blocks where $\B = \bm{0}$ the
ring CCD equation (Eq.  \ref{Riccati}) becomes
\begin{equation} 
\A \, \T + \T \, \A = \bm{0}. 
\label{eq:B0block}
\end{equation} 
Because the direct RPA $\A$ is positive definite,\cite{Scuseria2009} the only
solution is $\T=\bm{0}$. Thus the spin flip components of $\T$ vanish, and in
the closed shell case all triplet components of $\T$ vanish.  Triplets do not
contribute to direct RPA.

A third advantage of direct RPA is that the correlation energy expression is 
well defined, unlike in full RPA.\cite{Oddershede}  While in both direct and 
full RPA, we have 
$\mathrm{Tr}(\B \, \T) = \mathrm{Tr}(\w - \A)$,\cite{Scuseria2009} 
only in direct RPA is the prefactor in the correlation energy unambiguously 
1/2.  For full RPA, the plasmonic formula suggests the prefactor should be 
1/2, while the connection to CC theory suggests that the prefactor should be 
1/4.  Jeziorski and coworkers have presented another alternative full RPA 
energy expression that uses the plasmonic prefactor of 1/2, but subtracts the 
second-order MP2 correlation energy.\cite{Jeziorski1993}  

The elimination of exchange integrals in direct RPA may appear artificial from
a wave function perspective, and in fact the wave function underlying direct
RPA can violate the Pauli exclusion principle.  However, direct RPA can be
rigorously derived by applying the adiabatic connection fluctuation-dissipation
theorem to the Kohn-Sham noninteracting reference
system.\cite{Furche2001,Furche2005}  On balance, long-range direct RPA from the
GKS reference appears to be a practical ansatz for long-range correlation.

\begin{figure}[ht]
\includegraphics[width=0.47\textwidth]{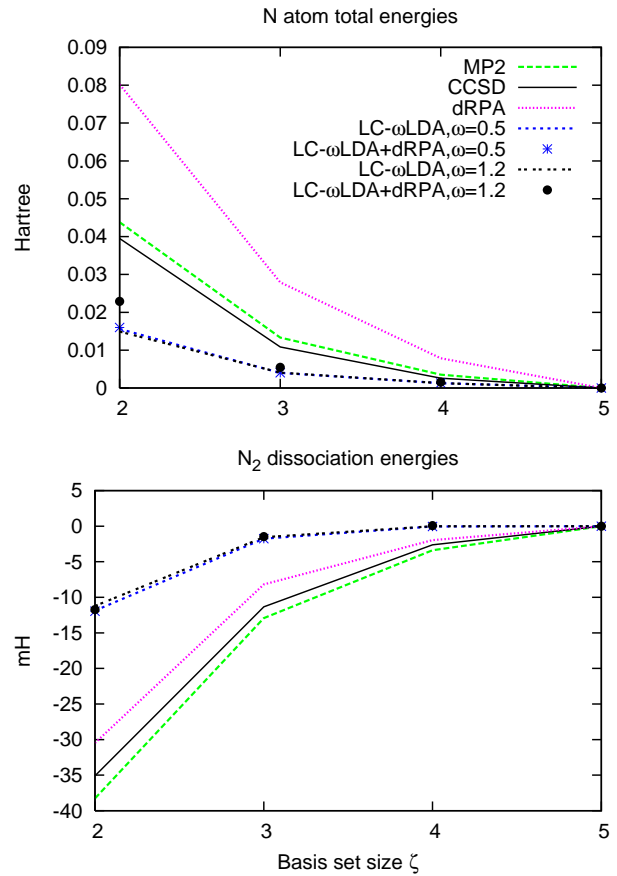}
\caption{\label{fig:bs} Basis set dependence of \LRdRPA.  (Top) Total energy 
of N atom.  (Bottom) Dissociation energy of N$_2$ at experimental bond length
1.098 Angstrom.  Results are calculated with aug-cc-pV$\zeta$Z basis sets, and
normalized to the aug-cc-pV5Z result. \LRdRPA\ calculations use $c_{RPA}=1$ 
and either $\omega=0.5$ or $\omega=1.2$ Bohr$^{-1}$.}
\end{figure}

{\em Computational details.} We have implemented our expressions into the
development version of the \GDV\ suite of programs.\cite{GDV-G1}  Matrices $\A$
and $\B$ are evaluated in the full spin-orbital basis set, without symmetry
adaptation.  We use the range-separated LSDA correlation functional of Paziani and
coworkers.\cite{Paziani2006}  In what follows, ``\LRHF'' denotes GKS
calculations using short-range LSDA xc and long-range HF exchange.  (Note that
in our previous papers, this acronym denoted SR LSDA exchange, LR HF exchange,
and \textit{full range} LSDA correlation.) Adding long-range direct RPA
correlation as described above results in ``\LRdRPA''.  The ``dRPA'' acronym by
itself  denotes conventional, full range
({\em i.e.}, not range separated) HF exchange and direct RPA correlation; this
``dRPA'' energy is evaluated from self-consistent Kohn-Sham orbitals and
orbital energies calculated with the Perdew-Burke-Ernzerhof (PBE) generalized
gradient (GGA) xc functional.\cite{pbe} Open-shell systems are
treated spin unrestricted.  Eq. \ref{Riccati} is solved iteratively using
DIIS\cite{Pulay1982,Scuseria1986} for coupled clusters.  Correlated
calculations use frozen core electrons.

Like other
workers,\cite{Stoll1985,rsh1,Leininger1997,Iikura2001,Heyd2003,Toulouse2004,Goll2005,Angyan,Goll2006,Gerber2007,Henderson2007}
we select the range separation parameter $\omega$ empirically.  Our standard
\LRdRPA\ calculations set $c_{RPA}=1$ in Eq. \ref{eq:Exctot}.  We also explore
treating $c_{RPA}$ as an empirical parameter.

\begin{figure}[t]
\includegraphics[width=0.47\textwidth]{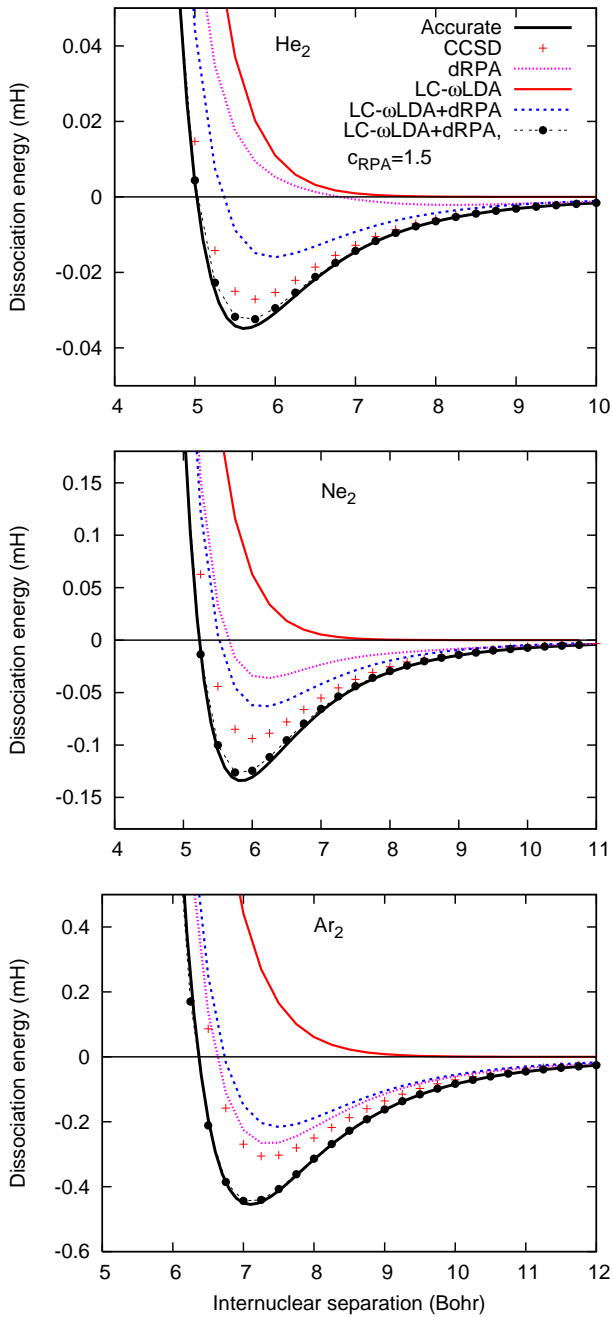}
\caption{\label{fig:vdW} Counterpoise-corrected dissociation curves of van der 
Waals dimers. Aug-cc-pV5Z basis, other details in the text.}
\end{figure}

\begin{table}[ht]
\caption{\label{tab:AE} Mean absolute errors (kcal/mol) in AE6 atomization
energies, G2/97 heats of formation, BH6 and HTBH38/04 hydrogen-transfer reaction
barrier heights, and NHTBH38/04 non-hydrogen-transfer barrier heights.
6-311+G(2d,2p) basis set. $\omega$ in Bohr$^{-1}$.}
\begin{tabular*}{0.5\textwidth}[c]{@{\extracolsep{\fill}}l r rr rrr  }
\hline\hline
Method&$\omega$&AE6&G2&BH6&HT&NHT\\
\hline
LC-$\omega$PBE\footnote{Ref. \onlinecite{LCwPBE}} 
         & 0.4 & 5.5 & 4.2 & 1.2 & 1.3 & 2.0 \\
\LRHF\footnote{Short-range LSDA exchange-correlation}    
         & 0.5 & 5.8 & 7.0 & 2.3 & 3.0 & 4.4 \\
\LRdRPA
         & 0.7 & 5.6 & 6.2 & 1.8 & 2.3 & 3.5 \\
\LRdRPA\footnote{c$_{RPA}=1.5$} 
         & 1.2 & 4.0 & 4.4 & 1.2 & 1.6 & 3.5 \\
\hline\hline
\end{tabular*}
\end{table}

\begin{figure}
\includegraphics[width=0.47\textwidth]{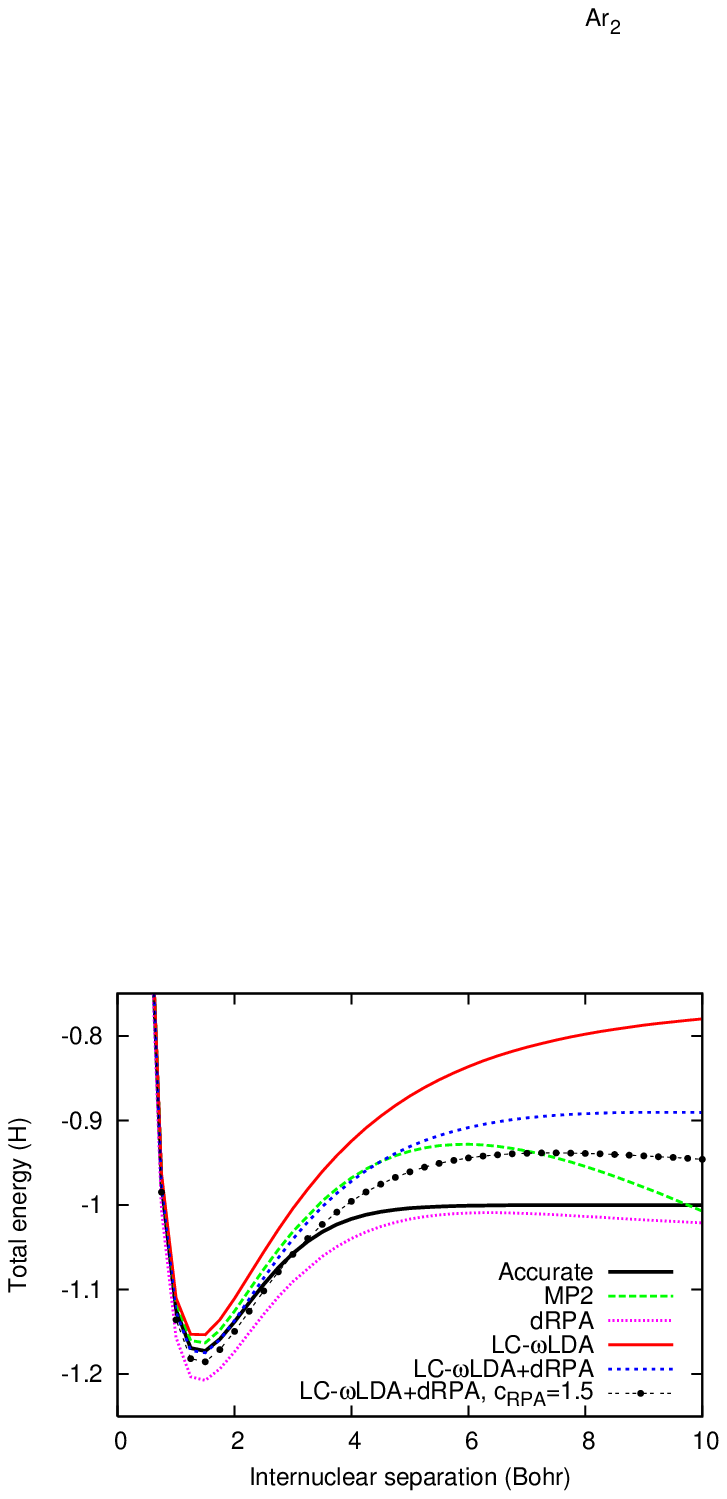}
\caption{\label{fig:H2} Dissociation of symmetry-restricted singlet H$_2$. Aug-cc-pVTZ 
basis set, $\omega=1.2$ Bohr$^{-1}$. "Accurate" results are 
full CI/aug-cc-pV5Z. }
\end{figure}

{\em Numerical results.} Figure \ref{fig:bs} shows the basis set dependence of
\LRdRPA\ for the total energy of N atom (top) and the dissociation energy of
N$_2$ (bottom).  As in previous work,\cite{Goll2006,Goll2008,Toulouse2009} the
long-range correlation has a much weaker basis set dependence than full-range
correlation.  Calculations at the relatively large $\omega=1.2$ Bohr$^{-1}$,
which incorporate a larger fraction of direct RPA correlation, converge more slowly
with basis set size.

Figure \ref{fig:vdW} shows counterpoise-corrected dissociation curves of van
der Waals dimers He$_2$, Ne$_2$, and Ar$_2$, evaluated in the large aug-cc-pV5Z
basis set.\cite{basis-sets}  Accurate curves are from Ref.
\onlinecite{TangToennies}.  \LRHF\ and \LRdRPA\ use $\omega=1.2$ Bohr$^{-1}$.
Rescaling the long-range dRPA correlation with $c_{RPA}=1.5$ significantly
improves the results, suggesting that the rescaling primarily corrects for
beyond-dRPA correlation effects rather than basis set incompleteness. The weak
He$_2$  binding of standard (full-range) dRPA is increased by using
BP86 rather than PBE orbitals and orbital energies, consistent with Ref.
\onlinecite{Furche2005} (not shown).

Table \ref{tab:AE} shows mean absolute errors  in the small AE6 and BH6 sets of
6 atomization energies and 6 reaction barrier heights,\cite{ae6bh6} the G2/97
set of 148 heats of formation\cite{G2} and the HTBH38/04 and NHTBH38/04 sets of
38 hydrogen-transfer and 38 non-hydrogen-transfer barrier
heights.\cite{BH42,NHTBH38} Calculations use the 6-311+G(2d,2p) basis set.
G2/97 calculations use B3LYP/6-31G(2df,p) geometries and vibrational
frequencies,\cite{G3geom} other geometries and reference values are taken from
Refs. \onlinecite{ae6bh6,G2,BH42,NHTBH38}. Results are presented for the
thermochemically optimal $\omega$.  \LRdRPA\ is quite accurate for
thermochemistry and kinetics, particularly with the empirical rescaling
$c_{RPA}=1.5$.  
It  improves upon the underlying \LRHF, giving results comparable to the accurate
LC-$\omega$PBE combination of range-separated GGA exchange and {\em full-range}
GGA correlation.\cite{LCwPBE} Adding long-range dRPA to LC-$\omega$PBE appears
to give significant double-counting of correlation (not shown).  However,
combining long-range dRPA with a short-range GGA (as in Ref.
\onlinecite{Toulouse2009}) may provide further improvements.

One of our goals is to apply range-separated dRPA to metallic systems.
This is possible because dRPA is robust as the band gap closes. Figure
\ref{fig:H2} illustrates \LRdRPA\ in one such prototypical system.  The figure
plots the energy of spin- and symmetry-restricted aug-cc-pVTZ H$_2$ as a
function of H-H bond length.  The MP2 correlation energy diverges as the bond
length increases and the HOMO-LUMO gap approaches zero.  Full RPA (not shown)
yields non-real correlation energies for bonds stretched beyond the
Coulson-Fischer point. In contrast, the \LRdRPA\ energy is real and finite at all bond
lengths.  Long-range dRPA does not capture all of the nondynamical correlation
present in stretched symmetric singlet H$_2$, and the dissociation limit is
thus predicted too high in energy. The functional also overestimates the energy
near equilibrium, especially with the empirical $c_{RPA}$ rescaling.  However,
there is a definite improvement over both \LRHF\ and MP2. 

The addition of long-range RPA correlation to range-separated hybrid density
functionals is a promising route to systematically incorporating nonlocal
correlation effects into DFT.  \LRdRPA\ can be evaluated in ${\cal O}(N^4)$
time via Cholesky decomposition of $\A$ and $\B$.  It is robust to unstable and
degenerate reference states, and shows promise for systems from metals, to
covalent bonds, to van der Waals complexes.

This work was supported by the National Science Foundation (CHE-0807194) and
the Welch Foundation (C-0036).  We thank Janos \'Angy\'an, Filipp Furche,
Andreas Savin, and Julien Toulouse for useful discussions.


\begin{thebibliography}{39}
\expandafter\ifx\csname natexlab\endcsname\relax\def\natexlab#1{#1}\fi
\expandafter\ifx\csname bibnamefont\endcsname\relax
  \def\bibnamefont#1{#1}\fi
\expandafter\ifx\csname bibfnamefont\endcsname\relax
  \def\bibfnamefont#1{#1}\fi
\expandafter\ifx\csname citenamefont\endcsname\relax
  \def\citenamefont#1{#1}\fi
\expandafter\ifx\csname url\endcsname\relax
  \def\url#1{\texttt{#1}}\fi
\expandafter\ifx\csname urlprefix\endcsname\relax\def\urlprefix{URL }\fi
\providecommand{\bibinfo}[2]{#2}
\providecommand{\eprint}[2][]{\url{#2}}

\bibitem[{\citenamefont{Oddershede}(1978)}]{Oddershede}
\bibinfo{author}{\bibfnamefont{J.}~\bibnamefont{Oddershede}},
  \bibinfo{journal}{Adv. Quant. Chem.} \textbf{\bibinfo{volume}{11}},
  \bibinfo{pages}{275} (\bibinfo{year}{1978}).

\bibitem[{\citenamefont{Ring and Schuck}(1980)}]{RingSchuck}
\bibinfo{author}{\bibfnamefont{P.}~\bibnamefont{Ring}} \bibnamefont{and}
  \bibinfo{author}{\bibfnamefont{P.}~\bibnamefont{Schuck}},
  \emph{\bibinfo{title}{The Nuclear Many-Body Problem}}
  (\bibinfo{publisher}{Springer-Verlag}, \bibinfo{address}{Berlin},
  \bibinfo{year}{1980}).

\bibitem[{\citenamefont{Furche}(2001)}]{Furche2001}
\bibinfo{author}{\bibfnamefont{F.}~\bibnamefont{Furche}},
  \bibinfo{journal}{Phys. Rev. B} \textbf{\bibinfo{volume}{64}},
  \bibinfo{pages}{195120} (\bibinfo{year}{2001}).

\bibitem[{\citenamefont{Dobson et~al.}(2005)\citenamefont{Dobson, Wang, Dinte,
  McLennan, and Le}}]{Dobson2005}
\bibinfo{author}{\bibfnamefont{J.~F.} \bibnamefont{Dobson}},
  \bibinfo{author}{\bibfnamefont{J.}~\bibnamefont{Wang}},
  \bibinfo{author}{\bibfnamefont{B.~P.} \bibnamefont{Dinte}},
  \bibinfo{author}{\bibfnamefont{K.}~\bibnamefont{McLennan}}, \bibnamefont{and}
  \bibinfo{author}{\bibfnamefont{H.~M.} \bibnamefont{Le}},
  \bibinfo{journal}{Int. J. Quant. Chem.} \textbf{\bibinfo{volume}{101}},
  \bibinfo{pages}{579} (\bibinfo{year}{2005}).

\bibitem[{\citenamefont{Furche and Van~Voorhis}(2005)}]{Furche2005}
\bibinfo{author}{\bibfnamefont{F.}~\bibnamefont{Furche}} \bibnamefont{and}
  \bibinfo{author}{\bibfnamefont{T.}~\bibnamefont{Van~Voorhis}},
  \bibinfo{journal}{J. Chem. Phys.} \textbf{\bibinfo{volume}{122}},
  \bibinfo{pages}{164106} (\bibinfo{year}{2005}).

\bibitem[{\citenamefont{Harl and Kresse}(2008)}]{Kresse}
\bibinfo{author}{\bibfnamefont{J.}~\bibnamefont{Harl}} \bibnamefont{and}
  \bibinfo{author}{\bibfnamefont{G.}~\bibnamefont{Kresse}},
  \bibinfo{journal}{Phys. Rev. B} \textbf{\bibinfo{volume}{77}},
  \bibinfo{pages}{045136} (\bibinfo{year}{2008}).

\bibitem[{\citenamefont{Furche}(2008)}]{Furche2008b}
\bibinfo{author}{\bibfnamefont{F.}~\bibnamefont{Furche}}, \bibinfo{journal}{J.
  Chem. Phys.} \textbf{\bibinfo{volume}{129}}, \bibinfo{pages}{114105}
  (\bibinfo{year}{2008}).

\bibitem[{\citenamefont{Scuseria et~al.}(2008)\citenamefont{Scuseria,
  Henderson, and Sorensen}}]{Scuseria2009}
\bibinfo{author}{\bibfnamefont{G.~E.} \bibnamefont{Scuseria}},
  \bibinfo{author}{\bibfnamefont{T.~M.} \bibnamefont{Henderson}},
  \bibnamefont{and} \bibinfo{author}{\bibfnamefont{D.~C.}
  \bibnamefont{Sorensen}}, \bibinfo{journal}{J. Chem. Phys.}
  \textbf{\bibinfo{volume}{129}}, \bibinfo{pages}{231101}
  (\bibinfo{year}{2008}).

\bibitem[{\citenamefont{Stoll and Savin}(1985)}]{Stoll1985}
\bibinfo{author}{\bibfnamefont{H.}~\bibnamefont{Stoll}} \bibnamefont{and}
  \bibinfo{author}{\bibfnamefont{A.}~\bibnamefont{Savin}}, in
  \emph{\bibinfo{booktitle}{Density Functional Methods in Physics}}, edited by
  \bibinfo{editor}{\bibfnamefont{R.}~\bibnamefont{Dreizler}} \bibnamefont{and}
  \bibinfo{editor}{\bibfnamefont{J.}~\bibnamefont{da~Providencia}}
  (\bibinfo{publisher}{Plenum}, \bibinfo{address}{New York},
  \bibinfo{year}{1985}), p. \bibinfo{pages}{177}.

\bibitem[{\citenamefont{Savin}(1996)}]{rsh1}
\bibinfo{author}{\bibfnamefont{A.}~\bibnamefont{Savin}}, in
  \emph{\bibinfo{booktitle}{Recent Developments and Applications of Modern
  Density Functional Theory}}, edited by \bibinfo{editor}{\bibfnamefont{J.~M.}
  \bibnamefont{Seminario}} (\bibinfo{publisher}{Elseveir},
  \bibinfo{address}{Amsterdam}, \bibinfo{year}{1996}), p. \bibinfo{pages}{327}.

\bibitem[{\citenamefont{Leininger et~al.}(1997)\citenamefont{Leininger, Stoll,
  Werner, and Savin}}]{Leininger1997}
\bibinfo{author}{\bibfnamefont{T.}~\bibnamefont{Leininger}},
  \bibinfo{author}{\bibfnamefont{H.}~\bibnamefont{Stoll}},
  \bibinfo{author}{\bibfnamefont{H.-J.} \bibnamefont{Werner}},
  \bibnamefont{and} \bibinfo{author}{\bibfnamefont{A.}~\bibnamefont{Savin}},
  \bibinfo{journal}{Chem. Phys. Lett.} \textbf{\bibinfo{volume}{275}},
  \bibinfo{pages}{151} (\bibinfo{year}{1997}).

\bibitem[{\citenamefont{Iikura et~al.}(2001)\citenamefont{Iikura, Tsuneda,
  Yanai, and Hirao}}]{Iikura2001}
\bibinfo{author}{\bibfnamefont{H.}~\bibnamefont{Iikura}},
  \bibinfo{author}{\bibfnamefont{T.}~\bibnamefont{Tsuneda}},
  \bibinfo{author}{\bibfnamefont{T.}~\bibnamefont{Yanai}}, \bibnamefont{and}
  \bibinfo{author}{\bibfnamefont{K.}~\bibnamefont{Hirao}}, \bibinfo{journal}{J.
  Chem. Phys.} \textbf{\bibinfo{volume}{115}}, \bibinfo{pages}{3540}
  (\bibinfo{year}{2001}).

\bibitem[{\citenamefont{Heyd et~al.}(2003)\citenamefont{Heyd, Scuseria, and
  Ernzerhof}}]{Heyd2003}
\bibinfo{author}{\bibfnamefont{J.}~\bibnamefont{Heyd}},
  \bibinfo{author}{\bibfnamefont{G.~E.} \bibnamefont{Scuseria}},
  \bibnamefont{and}
  \bibinfo{author}{\bibfnamefont{M.}~\bibnamefont{Ernzerhof}},
  \bibinfo{journal}{J. Chem. Phys.} \textbf{\bibinfo{volume}{118}},
  \bibinfo{pages}{8207} (\bibinfo{year}{2003}), \bibinfo{note}{{\bf{124}},
  219906(E) (2006)}.

\bibitem[{\citenamefont{Toulouse et~al.}(2004)\citenamefont{Toulouse, Colonna,
  and Savin}}]{Toulouse2004}
\bibinfo{author}{\bibfnamefont{J.}~\bibnamefont{Toulouse}},
  \bibinfo{author}{\bibfnamefont{F.}~\bibnamefont{Colonna}}, \bibnamefont{and}
  \bibinfo{author}{\bibfnamefont{A.}~\bibnamefont{Savin}},
  \bibinfo{journal}{Phys. Rev. A} \textbf{\bibinfo{volume}{70}},
  \bibinfo{pages}{062505} (\bibinfo{year}{2004}).

\bibitem[{\citenamefont{Goll et~al.}(2005)\citenamefont{Goll, Werner, and
  Stoll}}]{Goll2005}
\bibinfo{author}{\bibfnamefont{E.}~\bibnamefont{Goll}},
  \bibinfo{author}{\bibfnamefont{H.-J.} \bibnamefont{Werner}},
  \bibnamefont{and} \bibinfo{author}{\bibfnamefont{H.}~\bibnamefont{Stoll}},
  \bibinfo{journal}{Phys. Chem. Chem. Phys.} \textbf{\bibinfo{volume}{7}},
  \bibinfo{pages}{3917} (\bibinfo{year}{2005}).

\bibitem[{\citenamefont{\'Angy\'an et~al.}(2005)\citenamefont{\'Angy\'an,
  Gerber, Savin, and Toulouse}}]{Angyan}
\bibinfo{author}{\bibfnamefont{J.~C.} \bibnamefont{\'Angy\'an}},
  \bibinfo{author}{\bibfnamefont{I.~C.} \bibnamefont{Gerber}},
  \bibinfo{author}{\bibfnamefont{A.}~\bibnamefont{Savin}}, \bibnamefont{and}
  \bibinfo{author}{\bibfnamefont{J.}~\bibnamefont{Toulouse}},
  \bibinfo{journal}{Phys. Rev. A} \textbf{\bibinfo{volume}{72}},
  \bibinfo{pages}{012510} (\bibinfo{year}{2005}).

\bibitem[{\citenamefont{Goll et~al.}(2006)\citenamefont{Goll, Werner, Stoll,
  Leininger, Gori-Giorgi, and Savin}}]{Goll2006}
\bibinfo{author}{\bibfnamefont{E.}~\bibnamefont{Goll}},
  \bibinfo{author}{\bibfnamefont{H.-J.} \bibnamefont{Werner}},
  \bibinfo{author}{\bibfnamefont{H.}~\bibnamefont{Stoll}},
  \bibinfo{author}{\bibfnamefont{T.}~\bibnamefont{Leininger}},
  \bibinfo{author}{\bibfnamefont{P.}~\bibnamefont{Gori-Giorgi}},
  \bibnamefont{and} \bibinfo{author}{\bibfnamefont{A.}~\bibnamefont{Savin}},
  \bibinfo{journal}{Chem. Phys.} \textbf{\bibinfo{volume}{329}},
  \bibinfo{pages}{276} (\bibinfo{year}{2006}).

\bibitem[{\citenamefont{Gerber and \'Angy\'an}(2007)}]{Gerber2007}
\bibinfo{author}{\bibfnamefont{I.~C.} \bibnamefont{Gerber}} \bibnamefont{and}
  \bibinfo{author}{\bibfnamefont{J.~G.} \bibnamefont{\'Angy\'an}},
  \bibinfo{journal}{J. Chem. Phys.} \textbf{\bibinfo{volume}{126}},
  \bibinfo{pages}{044103} (\bibinfo{year}{2007}).

\bibitem[{\citenamefont{Henderson et~al.}(2007)\citenamefont{Henderson,
  Izmaylov, Scuseria, and Savin}}]{Henderson2007}
\bibinfo{author}{\bibfnamefont{T.~M.} \bibnamefont{Henderson}},
  \bibinfo{author}{\bibfnamefont{A.~F.} \bibnamefont{Izmaylov}},
  \bibinfo{author}{\bibfnamefont{G.~E.} \bibnamefont{Scuseria}},
  \bibnamefont{and} \bibinfo{author}{\bibfnamefont{A.}~\bibnamefont{Savin}},
  \bibinfo{journal}{J. Chem. Phys.} \textbf{\bibinfo{volume}{127}},
  \bibinfo{pages}{221103} (\bibinfo{year}{2007}).

\bibitem[{\citenamefont{Toulouse et~al.}()\citenamefont{Toulouse, Gerber,
  Jansen, Savin, and \'Angy\'an}}]{Toulouse2009}
\bibinfo{author}{\bibfnamefont{J.}~\bibnamefont{Toulouse}},
  \bibinfo{author}{\bibfnamefont{I.~C.} \bibnamefont{Gerber}},
  \bibinfo{author}{\bibfnamefont{G.}~\bibnamefont{Jansen}},
  \bibinfo{author}{\bibfnamefont{A.}~\bibnamefont{Savin}}, \bibnamefont{and}
  \bibinfo{author}{\bibfnamefont{J.~G.} \bibnamefont{\'Angy\'an}},
  \bibinfo{note}{{Phys. Rev. Lett.  in press, arXiv:0812.3302v2}}.

\bibitem[{\citenamefont{Kohn et~al.}(1998)\citenamefont{Kohn, Meir, and
  Makarov}}]{Kohn1998}
\bibinfo{author}{\bibfnamefont{W.}~\bibnamefont{Kohn}},
  \bibinfo{author}{\bibfnamefont{Y.}~\bibnamefont{Meir}}, \bibnamefont{and}
  \bibinfo{author}{\bibfnamefont{D.~E.} \bibnamefont{Makarov}},
  \bibinfo{journal}{Phys. Rev. Lett.} \textbf{\bibinfo{volume}{80}},
  \bibinfo{pages}{4153} (\bibinfo{year}{1998}).

\bibitem[{\citenamefont{Seidl et~al.}(1996)\citenamefont{Seidl, G\"orling,
  Vogl, Majewski, and Levy}}]{Seidl1996}
\bibinfo{author}{\bibfnamefont{A.}~\bibnamefont{Seidl}},
  \bibinfo{author}{\bibfnamefont{A.}~\bibnamefont{G\"orling}},
  \bibinfo{author}{\bibfnamefont{P.}~\bibnamefont{Vogl}},
  \bibinfo{author}{\bibfnamefont{J.~A.} \bibnamefont{Majewski}},
  \bibnamefont{and} \bibinfo{author}{\bibfnamefont{M.}~\bibnamefont{Levy}},
  \bibinfo{journal}{Phys. Rev. B} \textbf{\bibinfo{volume}{53}},
  \bibinfo{pages}{3764} (\bibinfo{year}{1996}).

\bibitem[{\citenamefont{Seeger and Pople}(1977)}]{Seeger1977}
\bibinfo{author}{\bibfnamefont{R.}~\bibnamefont{Seeger}} \bibnamefont{and}
  \bibinfo{author}{\bibfnamefont{J.~A.} \bibnamefont{Pople}},
  \bibinfo{journal}{J. Chem. Phys.} \textbf{\bibinfo{volume}{66}},
  \bibinfo{pages}{3045} (\bibinfo{year}{1977}).

\bibitem[{\citenamefont{Lepetit and Malrieu}(1990)}]{Malrieu}
\bibinfo{author}{\bibfnamefont{M.~B.} \bibnamefont{Lepetit}} \bibnamefont{and}
  \bibinfo{author}{\bibfnamefont{J.~P.} \bibnamefont{Malrieu}},
  \bibinfo{journal}{Chem. Phys. Lett.} \textbf{\bibinfo{volume}{169}},
  \bibinfo{pages}{285} (\bibinfo{year}{1990}).

\bibitem[{\citenamefont{Moszynski et~al.}(1993)\citenamefont{Moszynski,
  Jeziorski, and Szalewicz}}]{Jeziorski1993}
\bibinfo{author}{\bibfnamefont{R.}~\bibnamefont{Moszynski}},
  \bibinfo{author}{\bibfnamefont{B.}~\bibnamefont{Jeziorski}},
  \bibnamefont{and}
  \bibinfo{author}{\bibfnamefont{K.}~\bibnamefont{Szalewicz}},
  \bibinfo{journal}{Int. J. Quant. Chem.} \textbf{\bibinfo{volume}{45}},
  \bibinfo{pages}{409} (\bibinfo{year}{1993}).

\bibitem[{GDV()}]{GDV-G1}
\bibinfo{note}{Gaussian Development Version, Revision G.01, M. J. Frisch {\em
  et. al.}, Gaussian, Inc., Wallingford CT, 2007.}

\bibitem[{\citenamefont{Paziani et~al.}(2006)\citenamefont{Paziani, Moroni,
  Gori-Giorgi, and Bachelet}}]{Paziani2006}
\bibinfo{author}{\bibfnamefont{S.}~\bibnamefont{Paziani}},
  \bibinfo{author}{\bibfnamefont{S.}~\bibnamefont{Moroni}},
  \bibinfo{author}{\bibfnamefont{P.}~\bibnamefont{Gori-Giorgi}},
  \bibnamefont{and} \bibinfo{author}{\bibfnamefont{G.~B.}
  \bibnamefont{Bachelet}}, \bibinfo{journal}{Phys. Rev. B}
  \textbf{\bibinfo{volume}{73}}, \bibinfo{pages}{155111}
  (\bibinfo{year}{2006}).

\bibitem[{\citenamefont{Perdew et~al.}(1996)\citenamefont{Perdew, Burke, and
  Ernzerhof}}]{pbe}
\bibinfo{author}{\bibfnamefont{J.~P.} \bibnamefont{Perdew}},
  \bibinfo{author}{\bibfnamefont{K.}~\bibnamefont{Burke}}, \bibnamefont{and}
  \bibinfo{author}{\bibfnamefont{M.}~\bibnamefont{Ernzerhof}},
  \bibinfo{journal}{Phys. Rev. Lett.} \textbf{\bibinfo{volume}{77}},
  \bibinfo{pages}{3865} (\bibinfo{year}{1996}), \bibinfo{note}{{\bf{78}},
  1396(E) (1997)}.

\bibitem[{\citenamefont{Pulay}(1982)}]{Pulay1982}
\bibinfo{author}{\bibfnamefont{P.}~\bibnamefont{Pulay}}, \bibinfo{journal}{J.
  Comp. Chem.} \textbf{\bibinfo{volume}{3}}, \bibinfo{pages}{556}
  (\bibinfo{year}{1982}).

\bibitem[{\citenamefont{Scuseria et~al.}(1986)\citenamefont{Scuseria, Lee, and
  Schaefer~III}}]{Scuseria1986}
\bibinfo{author}{\bibfnamefont{G.~E.} \bibnamefont{Scuseria}},
  \bibinfo{author}{\bibfnamefont{T.~J.} \bibnamefont{Lee}}, \bibnamefont{and}
  \bibinfo{author}{\bibfnamefont{H.~F.} \bibnamefont{Schaefer~III}},
  \bibinfo{journal}{Chem. Phys. Lett.} \textbf{\bibinfo{volume}{130}},
  \bibinfo{pages}{236} (\bibinfo{year}{1986}).

\bibitem[{\citenamefont{Vydrov and Scuseria}(2006)}]{LCwPBE}
\bibinfo{author}{\bibfnamefont{O.~A.} \bibnamefont{Vydrov}} \bibnamefont{and}
  \bibinfo{author}{\bibfnamefont{G.~E.} \bibnamefont{Scuseria}},
  \bibinfo{journal}{J. Chem. Phys.} \textbf{\bibinfo{volume}{125}},
  \bibinfo{pages}{234109} (\bibinfo{year}{2006}).

\bibitem[{\citenamefont{Goll et~al.}(2008)\citenamefont{Goll, Werner, and
  Stoll}}]{Goll2008}
\bibinfo{author}{\bibfnamefont{E.}~\bibnamefont{Goll}},
  \bibinfo{author}{\bibfnamefont{H.-J.} \bibnamefont{Werner}},
  \bibnamefont{and} \bibinfo{author}{\bibfnamefont{H.}~\bibnamefont{Stoll}},
  \bibinfo{journal}{Chem. Phys.} \textbf{\bibinfo{volume}{346}},
  \bibinfo{pages}{257} (\bibinfo{year}{2008}).

\bibitem[{\citenamefont{Dunning~Jr.}(1989)}]{basis-sets}
\bibinfo{author}{\bibfnamefont{T.~H.} \bibnamefont{Dunning~Jr.}},
  \bibinfo{journal}{J. Chem. Phys.} \textbf{\bibinfo{volume}{90}},
  \bibinfo{pages}{1007} (\bibinfo{year}{1989}), \bibinfo{note}{{D. E. Woon and
  T. H. Dunning, Jr., J. Chem. Phys. {\bf 98}, 1358 (1993); {\bf 100}, 2975
  (1994)}}.

\bibitem[{\citenamefont{Tang and Toennies}(2003)}]{TangToennies}
\bibinfo{author}{\bibfnamefont{K.~T.} \bibnamefont{Tang}} \bibnamefont{and}
  \bibinfo{author}{\bibfnamefont{J.~P.} \bibnamefont{Toennies}},
  \bibinfo{journal}{J. Chem. Phys.} \textbf{\bibinfo{volume}{118}},
  \bibinfo{pages}{4976} (\bibinfo{year}{2003}).

\bibitem[{\citenamefont{Lynch and Truhlar}(2003)}]{ae6bh6}
\bibinfo{author}{\bibfnamefont{B.~J.} \bibnamefont{Lynch}} \bibnamefont{and}
  \bibinfo{author}{\bibfnamefont{D.~G.} \bibnamefont{Truhlar}},
  \bibinfo{journal}{J. Phys. Chem. A} \textbf{\bibinfo{volume}{107}},
  \bibinfo{pages}{8996} (\bibinfo{year}{2003}), \bibinfo{note}{{\bf{108}},
  1460(E) (2004)}.

\bibitem[{\citenamefont{Curtiss et~al.}(1997)\citenamefont{Curtiss,
  Raghavachari, Redfern, and Pople}}]{G2}
\bibinfo{author}{\bibfnamefont{L.~A.} \bibnamefont{Curtiss}},
  \bibinfo{author}{\bibfnamefont{K.}~\bibnamefont{Raghavachari}},
  \bibinfo{author}{\bibfnamefont{P.~C.} \bibnamefont{Redfern}},
  \bibnamefont{and} \bibinfo{author}{\bibfnamefont{J.~A.} \bibnamefont{Pople}},
  \bibinfo{journal}{J. Chem. Phys.} \textbf{\bibinfo{volume}{106}},
  \bibinfo{pages}{1063} (\bibinfo{year}{1997}).

\bibitem[{\citenamefont{Zhao et~al.}(2004)\citenamefont{Zhao, Lynch, and
  Truhlar}}]{BH42}
\bibinfo{author}{\bibfnamefont{Y.}~\bibnamefont{Zhao}},
  \bibinfo{author}{\bibfnamefont{B.~J.} \bibnamefont{Lynch}}, \bibnamefont{and}
  \bibinfo{author}{\bibfnamefont{D.~G.} \bibnamefont{Truhlar}},
  \bibinfo{journal}{J. Phys. Chem. A} \textbf{\bibinfo{volume}{108}},
  \bibinfo{pages}{2715} (\bibinfo{year}{2004}).

\bibitem[{\citenamefont{Zhao et~al.}(2005)\citenamefont{Zhao,
  Gonz\'ales-Garc\'ia, and Truhlar}}]{NHTBH38}
\bibinfo{author}{\bibfnamefont{Y.}~\bibnamefont{Zhao}},
  \bibinfo{author}{\bibfnamefont{N.}~\bibnamefont{Gonz\'ales-Garc\'ia}},
  \bibnamefont{and} \bibinfo{author}{\bibfnamefont{D.~G.}
  \bibnamefont{Truhlar}}, \bibinfo{journal}{J. Phys. Chem. A}
  \textbf{\bibinfo{volume}{109}}, \bibinfo{pages}{2012} (\bibinfo{year}{2005}),
  \bibinfo{note}{{\bf{110}}, 4942(E) (2006)}.

\bibitem[{\citenamefont{Curtiss et~al.}(2001)\citenamefont{Curtiss, Redfern,
  Raghavachari, and Pople}}]{G3geom}
\bibinfo{author}{\bibfnamefont{L.~A.} \bibnamefont{Curtiss}},
  \bibinfo{author}{\bibfnamefont{P.~C.} \bibnamefont{Redfern}},
  \bibinfo{author}{\bibfnamefont{K.}~\bibnamefont{Raghavachari}},
  \bibnamefont{and} \bibinfo{author}{\bibfnamefont{J.~A.} \bibnamefont{Pople}},
  \bibinfo{journal}{J. Chem. Phys.} \textbf{\bibinfo{volume}{114}},
  \bibinfo{pages}{108} (\bibinfo{year}{2001}).

\end{thebibliography}

\end{document}